# Probabilistic Symbolic-Distillation Model of Droplet Collision for Spray Simulation at High Ambient Pressures

Weiming Xu[1,2#], Tao Yang[1,2#], and Peng Zhang[1,2*]

*1. Department of Mechanical Engineering, City University of Hong Kong, Kowloon Tong, Kowloon, 999077, Hong Kong*
*2. Shenzhen Research Institute, City University of Hong Kong, Shenzhen, 518057, P. R. China*

**Abstract**

Droplet collision governs droplet population dynamics in many chemical engineering processes, such as spray drying, spray cooling, agricultural spraying, and combustion. Existing analytical models impose deterministic, pairwise boundaries between collision outcomes, whereas machine-learning classifiers lack the explicit functional form required of analytical collision submodels. In this study, we develop a probabilistic symbolic-distillation model using nearly forty thousand experimental events spanning eight regimes and five dimensionless parameters, including over five thousand data for ambient pressure up to 50 atm. A machine-learning teacher learns the joint outcome-probability landscape from these data, and symbolic regression subsequently distils it into eight class-specific expressions that jointly define a coupled analytical model. The resulting analytical field replaces abrupt regime switching with finite-width fuzzy boundaries. It outperforms the evaluated conventional analytical boundary models and reveals that their main limitation is the inability of zero-width boundaries to represent gradual probability transitions. The "biased-dice" sampling scheme provides a statistically consistent and practically convenient model implementation for Eulerian-Lagrangian spray simulation.

---

[*] Corresponding author
E-mail address: penzhang@cityu.edu.hk, Tel: (852)34429561.
[#] The authors equally contributed to the work.

## 1. Introduction

Sprays are central to a wide range of chemical and process engineering operations [1, 2], such as drying, cooling, agricultural spraying, and engine combustion. The droplet size distribution determines the process performance [2, 3], but the distribution is not fixed at atomisation. In dense spray regions, high droplet number density causes continuous collisions that redistribute mass and momentum among the droplets. Coalescence, separation after temporal coalescence, and splashing are the collision outcomes that affect droplet size distribution [2]. In the Eulerian-Lagrangian spray simulation, droplets are treated as discrete particles, and their collision partners are identified either deterministically or by stochastic sampling. The outcome of each identified collision is then set by a submodel [4-8]. Previous studies show that the choice of this outcome submodel alters the predicted spray evolution [3, 5, 9].

Binary droplet collisions are conventionally described by four dimensionless groups: Weber number ($We$), impact parameter ($B$), size ratio ($\Delta$), and Ohnesorge number ($Oh$). Outcomes are typically mapped on a $We - B$ nomogram containing four regimes: bouncing, coalescence, reflexive separation, and stretching separation, with the boundaries between them shifting with $\Delta$ and $Oh$ [10-13]. This four-regime division reflects the conditions accessible to early experiments. As experimental measurements extended in Weber number, size ratio, and the range of liquids examined, further outcomes were resolved: splashing and intense fragmentation at high $We$ [14, 15], rotational separation following temporary coalescence [16], and finger separation [11, 14].

Ambient gas pressure acts as another important parameter. It governs the drainage of the interstitial gas film between the droplets, which dictates whether their surfaces make contact. In general, higher pressure slows film drainage and promotes droplet bouncing, raising the critical Weber number required for coalescence [12, 17]. A recent experiment extending gauge pressure to 50 atm shows that the critical Weber number rises nonlinearly with pressure and then levels off above approximately 20 atm [18]. Therefore, existing collision models cannot predict the collision outcomes at high pressures and are applicable only within the parameter range covered by the data on which they were built.

In Eulerian-Lagrangian spray simulation, the treatment of droplet collision separates into three successive operations [2, 19-21]: detection of collision partners within the Lagrangian tracking loop, determination of the collision outcome from the local values of the governing parameters, and updating of the resulting droplet population in number, size, and velocity. The present work focuses on the second of these. The collision outcome model has conventionally been formulated as a composite of analytical criteria [20, 22]. Each criterion is a closed-form expression delimiting one pair of neighbouring outcomes in the $We - B$ plane: the lower bouncing boundary (LB), the reflexive separation–coalescence boundary (RS-C), and the stretching separation–coalescence boundary (SS-C), as shown in Fig. 1(a). The transition between soft coalescence and bouncing at very low Weber number is often ignored [21].

A composite model is assembled by selecting a formula for each transition boundary. By applying these formulas sequentially, every collision event is mapped to a specific region on the graph. For example, the construction of Munnannur and Reitz [23], combining the criteria of Estrade et al. [24], Ashgriz and Poo [10], and Brazier-Smith et al. [25], was implemented in KIVA [26]. In this way, evaluating this composite model based on a single aggregate accuracy can be misleading for two reasons. First, the transition criteria built for the model are independent. The overall performance is inherently constrained by its least accurate criterion. Second, an aggregate score averages all criteria together, weighted by how many data points fall in each regime. The same model tested on differently distributed data would score differently.

Some analytical criteria originate in energy analyses of the collision process. Specific boundaries were established by comparing kinetic and surface energies [10], rotational and surface energies [25], or by assessing the kinetic energy required to expel the interstitial gas film [24]. While physically intuitive, these derivations are inherently restricted, as they rely on highly idealized assumptions (e.g., inviscid flow, prescribed geometries) and were calibrated against narrow, atmospheric-pressure datasets. As broader measurements accumulated, subsequent models evolved primarily by empirically relaxing these initial restrictions. Researchers progressively incorporated viscous effects ($Oh$) and size-ratio ($\Delta$) dependencies into the classical boundaries for bouncing, reflexive, and stretching separation (e.g., Qian and Law [12], Gotaas et al. [27], Sommerfeld and Kuschel [13], and Sui et al. [28]). However, new data were fitted to extend the existing criteria rather than to revise them. Consequently, the validated parameter space grows incrementally, while ambient pressure and the non-classical outcomes have not entered these criteria, despite being well observed experimentally. Published evaluations demonstrate this predictive shortfall. For example, Agarwal et al. [29] tested the Munnannur and Reitz model, finding that its accuracy fell from 64% on early datasets to 43% on more recent ones [13]. These results indicate that empirical models are reliable only when the test conditions resemble the data they were calibrated on.

Alternatively, data-driven modeling infers collision outcomes directly from measurements. Agarwal [30] showed that machine-learning classifiers trained on 7898 experimental points exceeded 90% accuracy, significantly outperforming traditional physics-based models. Very recently, Yu and Chang [21] extended the training set to 30,809 points and coupled the machine-learning model with analytical criteria to revise conflicting data labels. While this approach achieved 94% training accuracy, performance fell to 82.2% on untrained data, which they attributed to measurement uncertainty in the collected experimental data.

Previous models, both analytical and data-driven, share a deterministic structure: each outputs a single regime label. Xu et al. [31] broke from this convention by treating collision outcomes as a probability distribution. Using a database of 33540 points, they trained a LightGBM (Light Gradient-Boosting Machine) [32] classifier across eight regimes using five parameters, including ambient pressure (up to 9 atm). Consequently, collision conditions are characterized by the relative likelihood of each outcome, allowing transitional regions to appear as continuous probability gradients rather than rigid boundaries. Because the tree-based classifier cannot be inspected or used outside the trained model, the probability field was projected onto a multinomial logistic regression in a second-order polynomial basis of the five parameters to recover an explicit form. A multinomial draw over the predicted probabilities then supplies the single definite outcome that Lagrangian tracking requires at each event. Performance was reported as aggregate accuracy, recall, and specificity over the full database.

For the droplet collision modelling, analytical criteria are limited by the data they were built on, while data-driven models are limited by their lack of interpretability. First, an analytical model is bound by its original assumptions and calibration range. Outside that range, it still gives a result, with no indication that its assumptions no longer hold. It works only in the regimes it was built to distinguish, and any other outcome is misclassified. It always represents boundaries as deterministic lines, while experiments show a probabilistic band. Second, current data-driven approaches capture these probabilistic transitions and provide usable formulas, but not interpretable ones. Their functional form is chosen based on simplified polynomials and only the coefficients are learned from the data, so the expression predicts well but reveals nothing about how the parameters act.

These two gaps converge on a single issue: how these models are evaluated. A composite model is made of separate criteria, one for each boundary. A new model should therefore be compared with it at each boundary separately, not by one overall score. Current data-driven models, however, report only a single accuracy over the whole database. Most of the data lie far from any boundary, so a model can score well overall while still getting the boundaries wrong. So far, no work has compared analytical criteria and data-driven models boundary by boundary, so which performs better there is still unknown. To address these gaps, we proposed a probabilistic symbolic-distillation model. The machine-learning model is not treated as the final predictor but as an intermediate representation from which analytical criteria are recovered. This separates two things that have so far been coupled: the model that fits the data need not be the model that is used, and the form of the criterion need not be assumed in advance.

A LightGBM teacher is trained on 38,762 experimental collision events spanning eight regimes and five parameters, with ambient pressure extending to 50 atm, a new record made very recently [18], and returns a probability distribution over all candidate outcomes rather than a single label, so that transitional bands are preserved as continuous variations rather than resolved into a boundary. Symbolic regression then distils this field into compact closed-form probability functions whose structure is found rather than assumed. Setting two of them equal yields an explicit transition boundary that can be written down and compared with the corresponding energy-based criterion, which is what separates a distilled criterion from a statistical surrogate, while a multinomial draw over the same expressions supplies the single definite outcome that Lagrangian tracking requires.

## 2. Machine-Learning Methodology

### 2.1 Experimental Database

The current database expands Yu and Chang [21] and our recent work [31] to 38,762 collision events by including the high-pressure measurements of Zhang et al. [18]. Each event is classified into one of eight collision regimes and characterized by five dimensionless parameters spanning broad ranges: $We = 0 \sim 2000$, $B = 0 \sim 1$, $\Delta = 1 \sim 5$, $Oh = 9.5 \times 10^{-4} \sim 5.5 \times 10^{-1}$, and $P = p/p_0 = 0.6 - 50$, where $p_0$ is the atmospheric pressure. The addition of over 5,000 high-pressure events is physically significant. Unlike earlier databases mostly restricted to ambient conditions, our previous database extended only to 9 atm but captured only the monotonic portion of the bouncing–coalescence transition. Consequently, models trained on such low-pressure data inherently over-extrapolate this trend, overestimating the critical Weber number at elevated pressures. The newly added data specifically span the pressure plateau [18].

Figure 1(a) displays the eight regimes in the $We$–$B$ plane with the LB, SS-C, and RS-C boundaries, while Fig. 1(b) shows their distributions in the present dataset: I, soft coalescence (274 points); II, bouncing (11,122 points); III, hard coalescence (14,715 points); IV, reflexive separation (2,879 points); V, stretching separation (9,175 points); VI, rotational separation (259 points); VII, finger separation (83 points); and VIII, splashing (255 points). Notably, the regimes are strongly imbalanced. Regimes II, III, and V comprise most of the data (up to 90%), while VI, VII, and VIII total under 600 events. This imbalance results from the measurement limits, instead of sampling bias. This inherent imbalance directly necessitates the specific learning algorithms and evaluation metrics, as adopted below.

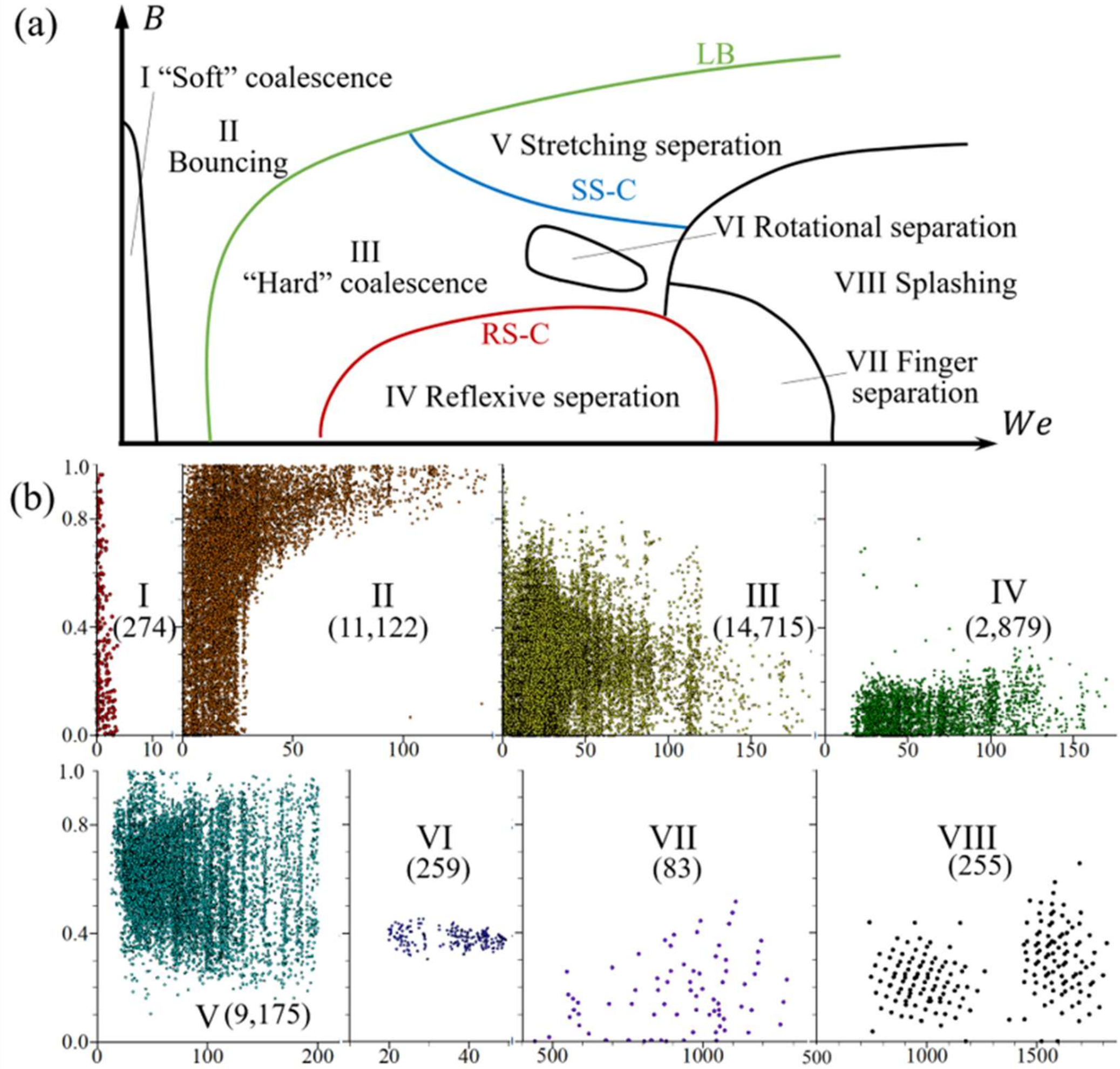

Fig. 1. The eight collision regimes in the $We - B$ plane. (a) Schematic nomogram showing the regimes and the three classical boundaries: the lower bouncing boundary (LB), the stretching separation–coalescence boundary (SS-C), and the reflexive separation–coalescence boundary (RS-C). (b) Distribution of the experimental data for each regime: soft coalescence after minor deformation (Regime I), bouncing (Regime II), hard coalescence after substantial deformation (Regime III), reflexive separation (Regime IV), stretching separation (Regime V), rotational separation (Regime VI), finger separation (Regime VII), and splashing (Regime VIII). The data point number in each regime is shown.

Figure 2 shows the pairwise distribution of the five parameters. Three features of the coverage bear on the modelling. First, the distribution of $We$ is multimodal, each mode corresponding to the range over which a group of regimes is observed. The parameters are therefore normalised before training. Second, the data of $Oh$, $\Delta$, and $P$ fall into discrete bands rather than filling the space, each band being one liquid, one size ratio, or one chamber pressure from a single study. The parameter space is thus sampled at points rather than continuously, and any model must interpolate between the bands. Third, $\Delta$ is concentrated near unity, as most reported experiments use equal or near-equal droplets. The collision events with $\Delta > 2$ are rare throughout. The elevated-pressure data are confined to regimes I–V, so the behaviour of regimes VI–VIII at $P > 1$ rests entirely on extrapolation.

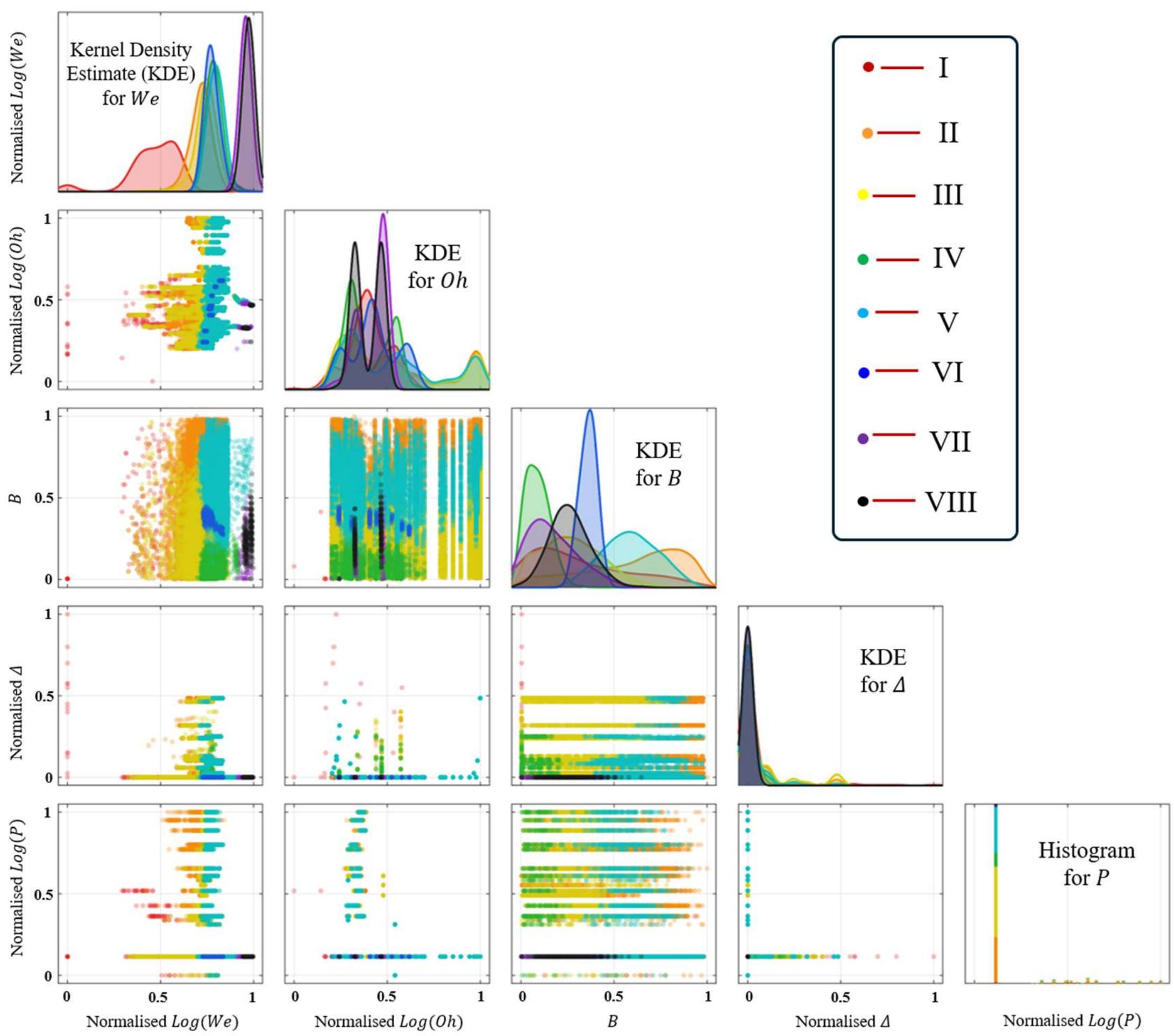

Fig. 2. Pairwise distribution of the five dimensionless parameters across the eight collision regimes. These off-diagonal panels show the joint distribution of each pair, while the diagonal panels show the kernel density estimate for each parameter (histogram for $P$). Colours denote the different regimes. $We$, $Oh$, and $P$ are plotted on logarithmic axes and all parameters are scaled to [0, 1].

These dataset characteristics dictate our modeling choices. First, because regime populations differ by up to three orders of magnitude, the learning algorithm must remain sensitive to rare outcomes represented by mere dozens of events. Consequently, model performance cannot be reliably evaluated via global aggregate accuracy. Second, since the parameter space is sampled in discrete bands rather than continuously, the model must handle sparse and irregular coverage robustly. Third, because nominally identical experiments within transitional bands often yield different outcomes, the learning target must be a probability distribution over candidate regimes rather than a single deterministic label.

The analytical distillation stage faces additional constraints. Specifically, the pressure dependence of the bouncing transition features a nonlinear saturation, and $We$ spans seven orders of magnitude. A pre-fixed polynomial basis cannot capture this saturation, and assuming a mathematical structure in advance risks enforcing behaviors unsupported by the data. Therefore, the optimal functional form must be discovered directly from the data.

**2.2 Overview of Probability Symbolic-distillation Method**

The probabilistic symbolic-distillation framework is summarized in Figure 3. Each collision event in the database shown in Fig. 3(a) is described by the input vector $\boldsymbol{x} = (P, We, B, \Delta, Oh)$ and its observed regime. These data are used to train the ensemble LightGBM model in Fig. 3(b), which estimates the normalized eight-regime probability vector $\boldsymbol{p}^T(\boldsymbol{x})$. The resulting probability fields assign comparable probabilities to neighbouring outcomes within model-inferred transition regions, while identifying a dominant outcome in well-separated regime interiors. These continuous probability fields subsequently serve as the targets for the symbolic-regression model in Fig. 3(c), using the regime-specific fitting strategies described in Section 2.4. The resulting expressions produce eight bounded class scores, which are aligned through class-specific affine transformations and jointly normalised by a softmax function to obtain the analytical probability vector $\boldsymbol{p}_k^{SR}(\boldsymbol{x})$. The expressions therefore operate as a coupled multiclass model rather than as independent binary transition criteria. Finally, when a simulation requires one outcome for an individual collision, the realised regime is drawn from the categorical distribution $Y \sim Cat[\boldsymbol{p}_k^{SR}(\boldsymbol{x})]$. This operation returns a single regime for each event while preserving the predicted probabilities in expectation. Across an ensemble of collisions, the resulting regime counts follow a multinomial distribution. The method thus connects the learned probability landscape, its explicit analytical representation, and the event-level outcome realization required for Eulerian-Lagrangian spray simulations.

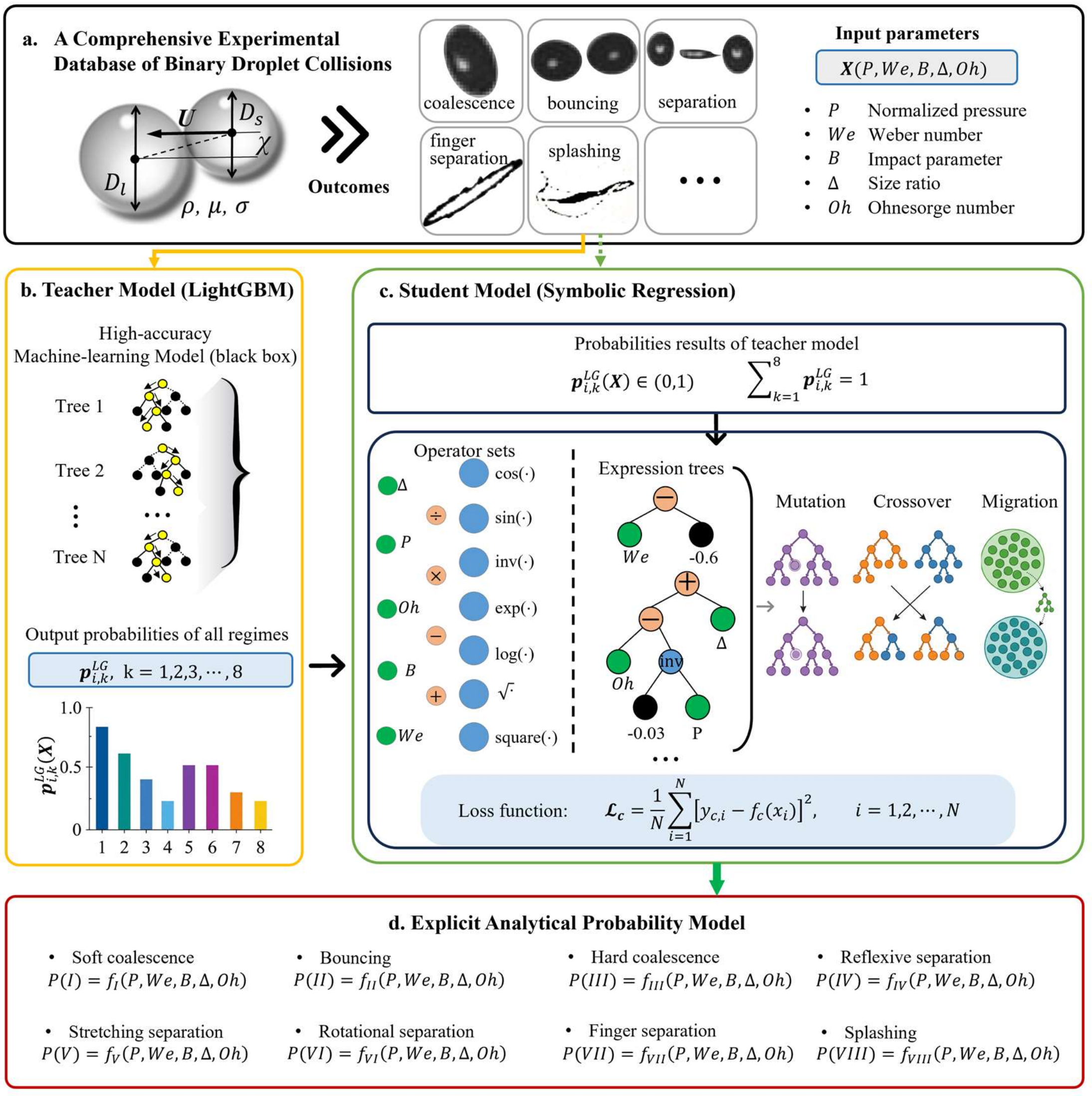

Fig. 3. Schematic of the probabilistic symbolic-distillation framework for droplet collision prediction, including (a) database input, (b) LightGBM machine learning model, (c) symbolic regression model, and (d) explicit analytical probability model.

### 2.3 LightGBM Teacher Model for Eight-regime Probability Learning

A reliable probabilistic teacher model is first constructed to approximate the regime distribution in parameter space with high accuracy and stability. In this framework, LightGBM is adopted, and droplet-collision prediction is formulated as an eight-regime probabilistic learning problem [32]. Each collision event is represented by the descriptor vector $\boldsymbol{x_i} = (P_i, We_i, B_i, \Delta_i, Oh_i)$, and the observed outcome $\boldsymbol{y_i} \in \{\boldsymbol{I}, \boldsymbol{II}, \dots, \boldsymbol{VIII}\}$ provides the training label. The teacher is thus trained to return an eight-component probability vector, thereby enabling a continuous representation of regime transitions rather than a single discrete assignment.

This algorithm is a histogram-based gradient-boosted decision-tree method well-suited for capturing nonlinear and conditional interactions among the collision descriptors. It was constructed as an ensemble of independently trained LightGBM models to reduce its sensitivity to the particular training sample. Each model was fitted to a separately resampled subset $\mathcal{D}_m$, containing 80% of the training events. For the $m$-th LightGBM estimator, the score assigned to regime $k$ after $T$ boosting iterations is written as

$$\boldsymbol{z}_{i,k}^{(m)} = \boldsymbol{z}_{i,k}^{(m,0)} + \eta \sum_{t=1}^{T} \boldsymbol{h}_{k,t}^{(m)}(\boldsymbol{x_i}) \tag{1}$$

where $\boldsymbol{z}_{i,k}^{(m,0)}$ is the initial score for regime $k$, $\boldsymbol{h}_{k,t}^{(m)}$ is the contribution of the tree added at iteration $t$, and $\eta$ is the learning rate. At every iteration, candidate splits are evaluated from the first- and second-order derivatives of the multiclass loss. The continuous predictors are binned into histograms, and the leaf with the largest loss reduction is expanded. This procedure permits the model to learn localized, conditional changes in regime occurrence without imposing a prescribed analytical switching boundary.

The accumulated class scores $\boldsymbol{z}_{i,k}^{(m)}$ quantify the relative support for the eight collision regimes rather than their individual occurrence probabilities. Because the regimes constitute mutually exclusive outcomes, each score must be evaluated relative to those of the competing regimes. The scores are therefore jointly normalised through the softmax function

$$\boldsymbol{q}_{i,k}^{(m)} = \frac{e^{\boldsymbol{z}_{i,k}^{(m)}}}{\sum_{j=1}^{8} e^{\boldsymbol{z}_{i,k}^{(m)}}}, k = \text{Ⅰ}, \cdots, \text{Ⅷ}, \sum_{k=1}^{8} \boldsymbol{q}_{i,k}^{(m)} = 1 \tag{2}$$

where $0 < \boldsymbol{q}_{i,k}^{(m)} < 1$ is the probability assigned to Regime *k*.

For each ensemble member, the tree structures and leaf values were determined by minimizing the class-weighted multiclass cross-entropy loss

$$\mathcal{L}^{(m)} = -\frac{1}{N_m} \sum_{i \in \mathcal{D}_m} w_{y_i} \log \boldsymbol{q}_{i,y_i}^{(m)} \tag{3}$$

where $N_m$ is the number of collision events. The weight assigned to Regime $k$ is $w_k = \frac{N}{K n_k}$, where $n_k$ is the number of events belonging to that regime in the development database. This weighting prevents the cross-entropy loss from being dominated by bouncing, hard coalescence, and stretching separation, while retaining the sparsely represented outcomes during teacher learning. The teacher consisted of $M = 10$ LightGBM estimators fitted to resampled subsets containing 80% of the training data. Each estimator used 200 boosting iterations, 31 leaves, a learning rate of 0.2, and a maximum tree depth of 10. The ensemble probability was obtained by averaging the outputs of the ten estimators,

$$\boldsymbol{p}_{i,k}^{LG} = \frac{1}{M} \sum_{m=1}^{10} \boldsymbol{q}_{i,k}^{(m)} \tag{4}$$

Each $\boldsymbol{q}_i^{(m)}$ is softmax-normalised, so the ensemble average also satisfies $\boldsymbol{0} < \boldsymbol{p}_{i,k}^{LG}(\boldsymbol{x}) < 1$, $\sum_{k=1}^{8} \boldsymbol{p}_{i,k}^{LG} = 1$. A deterministic teacher prediction may be recovered through $\hat{y}^{LG} = \arg\max_k \boldsymbol{p}_{i,k}^{LG}$. In the present framework, however, the complete probability vector $\boldsymbol{p}_i^{LG} = (\boldsymbol{p}_{i,1}^{LG}, \cdots, \boldsymbol{p}_{i,k}^{LG})$ is retained. A sharply concentrated vector identifies a condition for which the teacher assigns one dominant outcome, whereas comparable probabilities indicate reduced separability among neighbouring regimes. These continuous class-probability fields, rather than the discrete argmax labels, constitute the targets for the symbolic-distillation procedure described in Section 2.4.

### 2.4 Symbolic Regression Distillation of Probability Functions

Although the LightGBM teacher provides a full probability vector for each collision condition, deploying its tree ensemble directly in spray simulation is impractical due to computational cost and its black-box nature. We therefore employ symbolic regression to distill each component of the learned probability distribution into an explicit, closed-form expression composed of elementary mathematical functions [33]. The LightGBM ensemble provides, for each collision event, a teacher probability vector $\boldsymbol{p}_i^{LG} = (\boldsymbol{p}_{i,1}^{LG}, \cdots, \boldsymbol{p}_{i,k}^{LG})$. Symbolic regression (SR) was used to distil these continuous outputs rather than the discrete class labels, thereby retaining information on the relative likelihoods of competing outcomes.

A separate analytical representation was constructed for each regime using the five dimensionless descriptors $\boldsymbol{x_i} = (P_i, We_i, B_i, \Delta_i, Oh_i)$ . Because the teacher probabilities exhibit markedly different distributions among the eight regimes, a uniform regression strategy would be dominated by the numerous near-zero values and would inadequately resolve rare or localised outcomes.

Candidate expressions were evolved from the input variables using the basic arithmetic ($+$, $-$, $\times$, and $\div$), trigonometric ($sin$ and $cos$), exponential ($exp$), logarithmic ($ln$), and square root (√) operators. The symbolic regression was applied independently to each regime, so the fitting target was adapted to the distribution of the teacher probabilities.

Each teacher probability field $\boldsymbol{p}_i^{LG}$ was distilled into an analytical representation by symbolic regression. Because the eight fields differ substantially in both dynamic range and spatial support, fitting the raw probabilities uniformly would be inappropriate. In particular, for globally sparse regimes, the abundant near-zero values would dominate the regression objective and obscure the small but relevant variations that distinguish active from inactive regions. The regression target was therefore defined on a regime-specific basis, and the resulting expression was subsequently mapped to an intermediate regime score $\boldsymbol{r}_k$ through a fixed post-processing operation.

Three fitting strategies were adopted. For Regimes II–V, $\boldsymbol{p}_i^{LG}$ are broadly distributed across the unit interval within their active regions, providing sufficiently resolved targets for direct fitting. For Regimes I, VI, VII, and VIII, by contrast, the probability is concentrated near zero for most samples. For Regimes I and VIII, sparsity arises mainly from broadly distributed low probabilities rather than localised support. Symbolic regression was therefore applied to the log-odds of $\ln\frac{p_k}{1-p_k}$. For Regimes VI and VII, appreciable probabilities are spatially localised. Each field was therefore factorised into a localising component $f_k^g$, fitted over all samples to identify the active region, and a conditional-magnitude component $f_k^m$, fitted to the log-odds of $\boldsymbol{p}_i^{LG}$ using only the active samples.

In each case, the symbolic search minimised the weighted mean-square error between the expression output and its corresponding teacher-derived target, with fitting error and expression complexity treated as competing objectives. The final expression for each regime was selected from the Pareto front [34] based on fidelity to the teacher output, structural compactness, and numerical stability. The selected expressions were converted into bounded class scores according to

$$\boldsymbol{r}_k(\boldsymbol{x}) = \begin{cases} \sigma[f_k(\boldsymbol{x})], & k \in \{\text{ I , VIII}\} \\ \Pi_{[0,1]}[f_k(\boldsymbol{x})], & k \in \{\text{ II , III, IV, V }\} \\ \Pi_{[0,1]}\left[f_k^g(\boldsymbol{x})\right]\sigma[f_k^m(\boldsymbol{x})], & k \in \{\text{VI, VII}\} \end{cases} \tag{5}$$

where $\Pi_{[0,1]}(u) = \min(\max(u, 0), 1)$ denotes clipping to $[0,1]$. Accordingly, the raw-probability fits for Regimes II–V are clipped directly. For Regimes I and VIII, the expression lies on the log-odds scale and is mapped back to a probability by the sigmoid $\sigma$. Regimes VI and VII combine both mappings: the localising factor $f_k^g$ is clipped to $[0,1]$ to provide a continuous multiplicative gate, whereas the magnitude factor $f_k^m$, fitted on the log-odds scale, is transformed by the sigmoid to recover the probability magnitude within the active region.

These independently fitted outputs are bounded class scores, but they are not yet mutually normalised probabilities and do not generally satisfy $\sum_k r_k = 1$. To place the eight scores on a common scale, each $r_k$ was subjected to a class-specific affine transformation, followed by a joint softmax

$$\boldsymbol{p}_k^{SR}(\boldsymbol{x}) = \frac{e^{[\alpha_k \boldsymbol{r}_k(x)+\beta_k]}}{\sum_{j=1}^{K} e^{[\alpha_j \boldsymbol{r}_j(x)+\beta_j]}}, k = \text{I}, \cdots, \text{VIII}, \sum\nolimits_{k=1}^{k} \boldsymbol{p}_k^{SR}(\boldsymbol{x}) = 1 \quad (6)$$

The coupling coefficients $\{\alpha_k, \beta_k\}$ were determined jointly on a stratified calibration subset of the collected database by minimising the class-weighted multiclass cross-entropy against the observed collision labels.

## 2.5 Stochastic Realization of Symbolic Probability Outputs for E-L Simulation

Eulerian-Lagrangian spray simulations require each detected collision to be assigned a single outcome so that the corresponding regime-specific post-collision model can be activated [35]. By contrast, the symbolic model returns the complete probability vector $\boldsymbol{p}_k^{SR}(\boldsymbol{x})$. This vector is therefore not directly usable in such simulations. To resolve this, this probability distribution is converted into an event-level outcome through categorical sampling ("biased-dice" sampling [31]), as illustrated in Fig. 4. For a collision characterised by $\boldsymbol{x}$, a uniformly distributed random number $U \in [0,1)$ is generated, and the realised regime is determined from the cumulative probabilities

$$\hat{Y}(\boldsymbol{x}) = \min\{k: U < \sum_{j=1}^{k} \boldsymbol{p}_j^{SR}(\boldsymbol{x})\}, \; U \sim \mathcal{U}(0,1) \quad (7)$$

This construction ensures that the conditional probability of selecting regime $k$ is exactly $\boldsymbol{p}_k^{SR}(\boldsymbol{x})$. For a collection of collision events with different local conditions, the expected number of occurrences of regime $k$ is therefore $\sum_i \boldsymbol{p}_{i,k}^{SR}$. Event-level categorical draws consequently produce multinomial regime counts at the ensemble level, preserving the outcome distribution represented by the symbolic probability field. This procedure corresponds to the "biased random sampling", where the probability weights determine the likelihood of each face of the conceptual dice [36].

The stochastic realization has a clear limiting behaviour. When one probability approaches unity, the sampled outcome becomes effectively deterministic and coincides almost invariably with the argmax prediction. Within model-inferred transition regions, several neighbouring regimes retain appreciable probabilities and remain accessible under similar collision conditions. Sampling therefore retains this competition instead of collapsing the probability vector prematurely to a single deterministic boundary. This interpretation does not require every sampled event to reproduce its experimentally observed label. Its purpose in forward simulation is to preserve the regime frequencies implied by the probabilistic model. The argmax rule is retained separately for evaluating deterministic classification performance.

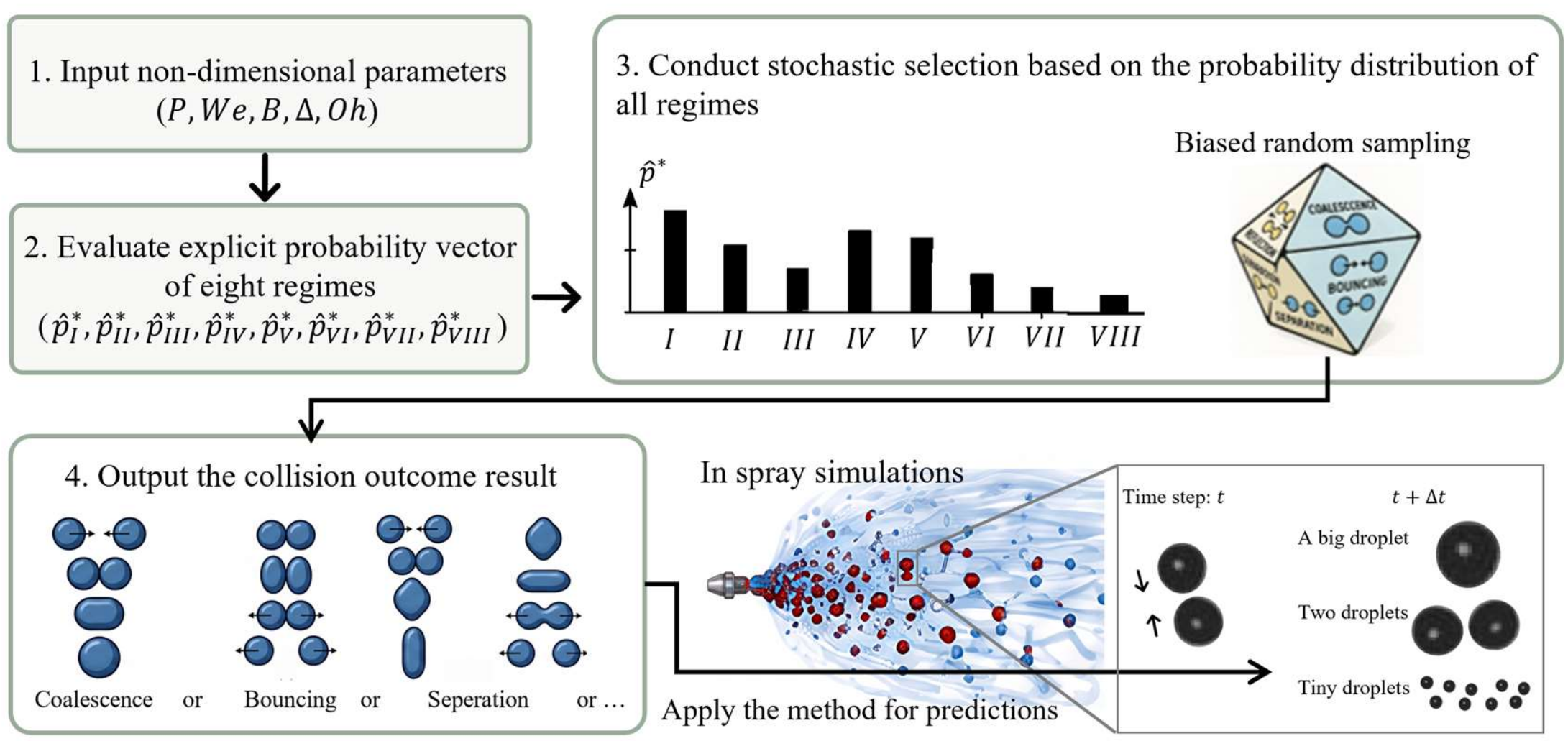

Fig. 4. Schematic of the stochastic regime selection from eight-class symbolic regression probability outcomes using multinomial distribution sampling.

In a spray simulation, the procedure is applied after a collision pair has been identified: the five dimensionless descriptors are evaluated from the pre-collision states, the eight analytical probabilities are computed, and one regime is sampled to activate the associated post-collision model. It therefore modifies only the outcome-selection stage and does not alter the collision-frequency or collision-partner models. Because the probability evaluation involves only closed-form expressions followed by a single random draw, the scheme can be incorporated without retaining the original LightGBM model or introducing an iterative inference procedure. Its distributional consistency, convergence with repeated realizations, and relation to deterministic argmax selection are examined in Section 3.3.

## 3. Results and Discussion

### 3.1 Closed-form Symbolic Probability Functions and Predictive Performance

Symbolic distillation constitutes the central step through which the learned collision-regime probability landscape is converted into an explicit analytical model. As listed in Table 1, the expressions therefore do not act as eight binary transition criteria. When evaluated together, these expressions provide the class scores underlying the coupled probability model. This section examines the predictive structure encoded by these expressions, their consistency with established physical behaviour, their performance over the collected database, and the efficiency of their analytical representation.

Table 1 shows that the eight collision outcomes retain distinct combinations of governing descriptors. The bouncing expression involves $We$ and $B$ with $Oh$ and $P$, whereas hard coalescence combines $We$, $B$, $\Delta$, and $Oh$. Reflexive separation is associated with sufficient inertia and near-head-on impact, while stretching separation is governed more strongly by the interaction between $We$ and $B$. Ambient pressure appears selectively rather than uniformly across the expressions.

The pressure dependence of the bouncing–coalescence transition provides an illustrative example of the physical trends recovered within the probabilistic representation. At fixed $\Delta$ and $Oh$, Fig. 5 extracts, for each $P$ and $B$, the critical Weber number $We_{cr}$ at the midpoint of the fuzzy transition and compares the resulting contour with the experimental data. Rather than being fitted separately as a deterministic boundary, this contour emerges directly from the joint probability field; while its midpoint enables comparison with conventional criteria, the underlying model retains the finite-width redistribution of probability among competing outcomes across the transition. The predicted $We_{cr}$ initially increases nonlinearly with pressure and subsequently approaches a plateau, in agreement with the experimental trend reported by Zhang et al. (2026) [18]. The coupled symbolic model can represent this behaviour because pressure is retained explicitly in the relevant class expressions, whereas conventional pressure-independent criteria based only on $We$ and $B$ necessarily predict an invariant boundary.

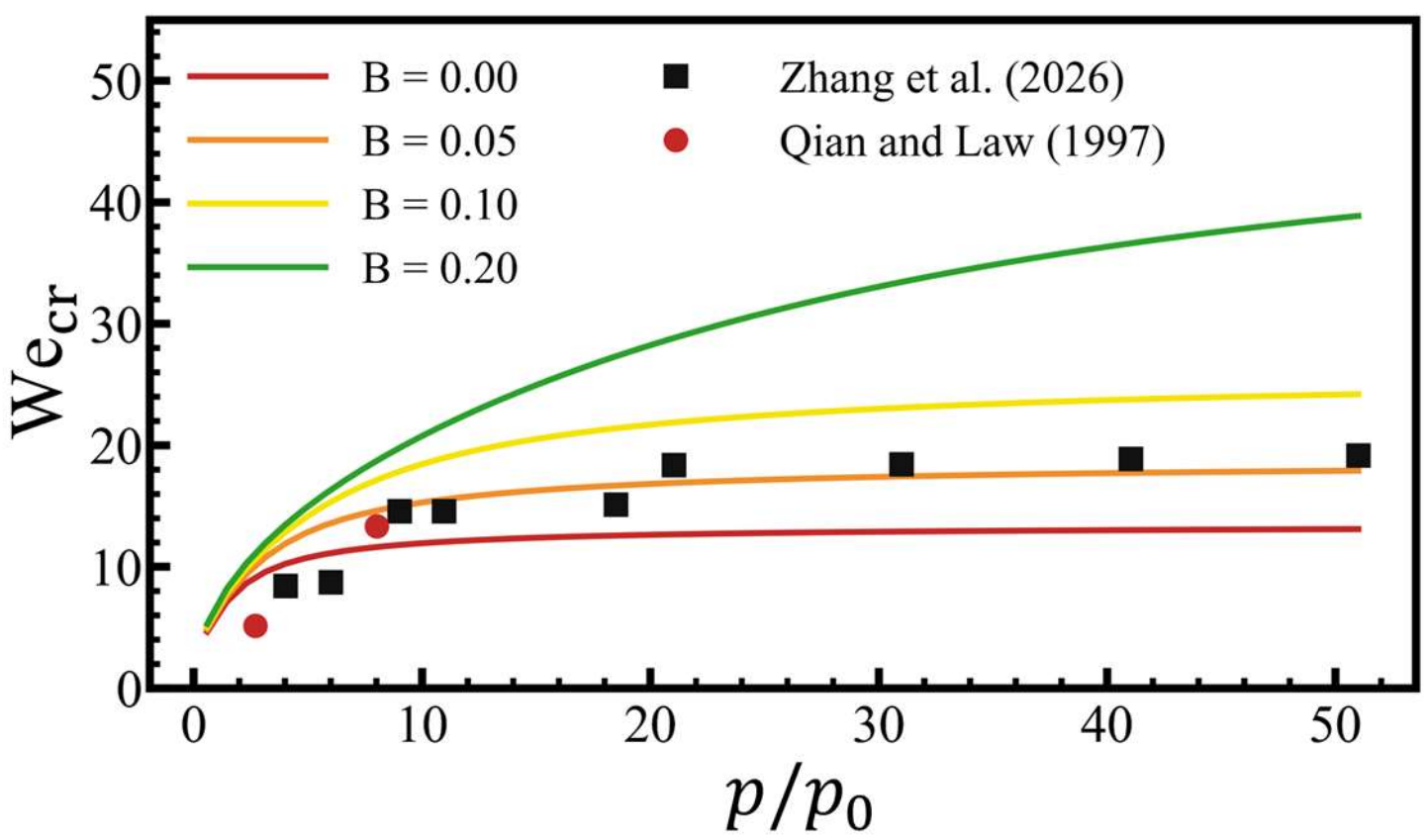

Fig. 5. Critical Weber number $We_{cr}$ at the midpoint of the bouncing–coalescence transition predicted by the SR model as a function of the pressure ratio $P = p/p_0$ for different impact parameters B, compared with experimental data (Qian and Law, 1997 [12]; Zhang et al., 2026 [18]). $\Delta = 1$, and $Oh = 0.08$.

Three representative expressions illustrate this physical consistency. For soft coalescence, the symbolic score decreases monotonically with $We$, and the associated probability changes from nearly unity below $We = 1$ to negligible values above $We = 5$. Consistently, all 274 soft-coalescence events occur below $We = 5$. The splashing expression contains only $We$ and $Oh$, consistent with the variable combination appearing in the classical Mundo–Cossali-type splashing parameter, $K \propto We^{0.625} Oh^{-0.25}$ [37], although not its exact threshold. This compact expression achieves a balanced accuracy of 0.998 and an AUROC of 1 in Figs. 6(a) and 6(b). For reflexive separation, the normalised sensitivity analysis assigns $B$ a dominant contribution of 0.905, while the exponential dependence on $B$ suppresses the score for increasingly oblique collisions and renders it negligible for $B > 0.5$. This agrees with the established confinement of reflexive separation to predominantly near-head-on collisions [10], where temporary coalescence is followed by strong axial deformation and recoil. Taken together, these examples show that the distilled expressions retain physically recognisable limits and variable combinations without imposing them beforehand.

1

Table 1. Symbolic regression analytical expressions for each collision regime, together with the softmax coupling coefficients $\{\alpha_k, \beta_k\}$.

| Regime | Formulas ($\boldsymbol{p}_k^{SR}(\boldsymbol{x}) = \frac{e^{[\alpha_k r_k(x)+\beta_k]}}{\sum_{j=1}^{K} e^{[\alpha_j r_j(x)+\beta_j]}}$, $x_0: P$, $x_1: We$, $x_2: B$, $x_3: \Delta$, and $x_4: Oh$) |
|---|---|
| **I** (Soft coalescence) | $\boldsymbol{r_0} = \boldsymbol{\sigma}(\boldsymbol{f_0}) = \frac{\boldsymbol{1}}{\boldsymbol{1+e^{-f_0}}}$, $\alpha_k = 10.59$, and $\beta_k = 7.05$; <br> $f_0 = 4.29\ \ln[\exp(2.49 - (\cos(x_1) + x_1) - \sin[2.18\sin(u)])) + 0.118]$, where $u = (-14.5\sqrt{x_4} + x_3) \cdot \cos\left(\frac{x_2}{-0.137}\right) + 2x_2$. |
| **II** (Bouncing) | $\boldsymbol{r_1} = \boldsymbol{clip}(\boldsymbol{f_1}, \boldsymbol{0}, \boldsymbol{1})$, $\alpha_k = 6.63$, and $\beta_k = 6.87$; <br> $f_1 = \sin\left[\exp\left(x_1 \cdot (1 - x_2)^2 \cdot \frac{-0.00131 x_1 \cdot {x_4}^{-0.4}}{x_0}\right)\right]$. |
| **III** (Hard coalescence) | $\boldsymbol{r_2} = \boldsymbol{clip}(\boldsymbol{f_2}, \boldsymbol{0}, \boldsymbol{1})$, $\alpha_k = 5.49$, and $\beta_k = 7.04$; <br> $f_2 = cos\left[sin(x_2 + \left[\frac{cos(0.00742 x_1 + (x_2/\sqrt{x_3})/0.391)}{-0.675}\right] - x_4) + 0.550\right] \cdot sin(\sqrt{x_3})$. |
| **IV** (Reflexive separation) | $\boldsymbol{r_3} = \boldsymbol{clip}(\boldsymbol{f_3}, \boldsymbol{0}, \boldsymbol{1})$, $\alpha_k = 8.54$, and $\beta_k = 7.00$; $f_3 = sin\left[\exp(1.10 - \frac{(2.19/x_1 + x_4) \cdot 9.04)}{exp(x_2/(-0.158))}\right]$. |
| **V** (Stretching separation) | $\boldsymbol{r_4} = \boldsymbol{clip}(\boldsymbol{f_4}, \boldsymbol{0}, \boldsymbol{1})$, $\alpha_k = 6.03$, and $\beta_k = 7.00$; $f_4 = sin[exp(sin[sin(-2.78 x_2) \cdot ln(x_1)]) \cdot x_2]$. |
| **VI** (Rotational separation) | $\boldsymbol{r_5} = \boldsymbol{clip}(f_5^g, \boldsymbol{0}, \boldsymbol{1}) \times \boldsymbol{\sigma}(f_5^m)$, $\alpha_k = 6.62$, and $\beta_k = 8.28$; <br> $f_5^g = cos[cos(cos(\frac{ln(x_1) \cdot x_2}{0.438}) \cdot (\frac{cos(x_2 \cdot 8.74) - 0.447)}{(x_4 - x_0 \cdot (-0.439)) \cdot x_3}) - 0.767] \cdot (-9.36)$, <br> $f_5^m = 2.19 \cdot \sin\left[\frac{1}{0.515} \sin\left(\sin \frac{\exp(sin\left(\frac{1.49}{\ln(cos(x_4))}\right) \cdot x_2)}{0.383 \big/ \sin\left(\frac{\ln x_1}{-0.816}\right)}\right) - 0.334\right]$. |
| **VII** (Finger separation) | $\boldsymbol{r_6} = \boldsymbol{clip}(f_6^g, \boldsymbol{0}, \boldsymbol{1}) \times \boldsymbol{\sigma}(f_6^m)$, $\alpha_k = 16.9$, and $\beta_k = 6.31$; <br> $f_6^g = sin[e^{x_2} \cdot \frac{0.404}{\cos((cos(ln(x_1 + 12.3) - 0.674) - 0.878) \cdot 2.35) - 0.762} - 0.974] \cdot 5.81 - 3.45$, <br> $f_6^m = sin[\frac{x_0 + x_2}{cos(x_2) + x_2 \cdot cos(x_1)} \cdot \frac{1}{sin({x_1}^2 \cdot 1.38 \cdot 10^{-6})}] + 0.890$. |
| **VIII** (Splashing) | $\boldsymbol{r_7} = \boldsymbol{\sigma}(\boldsymbol{f_7}) = \frac{\boldsymbol{1}}{\boldsymbol{1+e^{-f_7}}}$, $\alpha_k = 26.2$, and $\beta_k = 5.02$; <br> $f_7 = \frac{\sin\left[\ln\left(2.01 + \exp\left[\frac{sin(cos\left(\frac{-0.335}{x_4}\right) \cdot 0.00218 \cdot x_1 + 1.47)}{0.205}\right] + \exp\left[2.99 + x_1 \cdot sin(\frac{0.419}{x_4}) \cdot 0.00226\right]\right)\right]}{0.171} - 3.58$. |

2

On the collected eight-regime database, the symbolic model achieves a macro-averaged balanced accuracy of 0.918 under the argmax decision rule and a macro AUROC of 0.953, as shown in Figs. 6(a) and 6(b). The class-wise results show that predictive performance is closely related to regime separability. Soft coalescence (I) and finger separation (VII), which occupy comparatively distinct regions of the collision-parameter space, achieve balanced accuracies of 0.979 and 0.999, respectively. By contrast, bouncing (II) and hard coalescence (III) attain lower values of 0.834 and 0.775 because their predicted probabilities are often comparable within the bouncing–coalescence transition band. The representative $We - B$ map in Fig. 6(c) at $P = 1$, $\Delta = 1$, and $Oh = 0.01$, connects these performance differences to the spatial probability structure. A single outcome dominates within well-separated regime interiors, while neighbouring outcomes remain simultaneously probable across model-inferred finite-width transition regions. The symbolic model therefore retains strong probability-ranking capability while representing the reduced separability of competing outcomes near their transitions.

Figure 7(a) provides a complementary check of the class-wise results: the confusion matrix is largely diagonal, with the largest off-diagonal entries occurring between Regimes II and III, consistent with their comparatively lower balanced accuracies. Figure 7(b) shows that the symbolic model achieves a favourable accuracy–complexity balance. Increasing the polynomial order of the fixed-basis multinomial logistic-regression models substantially increases expression complexity, whereas the corresponding improvement in balanced accuracy is marginal. The symbolic model achieves higher accuracy than the higher-order polynomial models with a more compact expression because it assigns a separate nonlinear structure to each collision regime instead of expanding a common basis across all regimes.

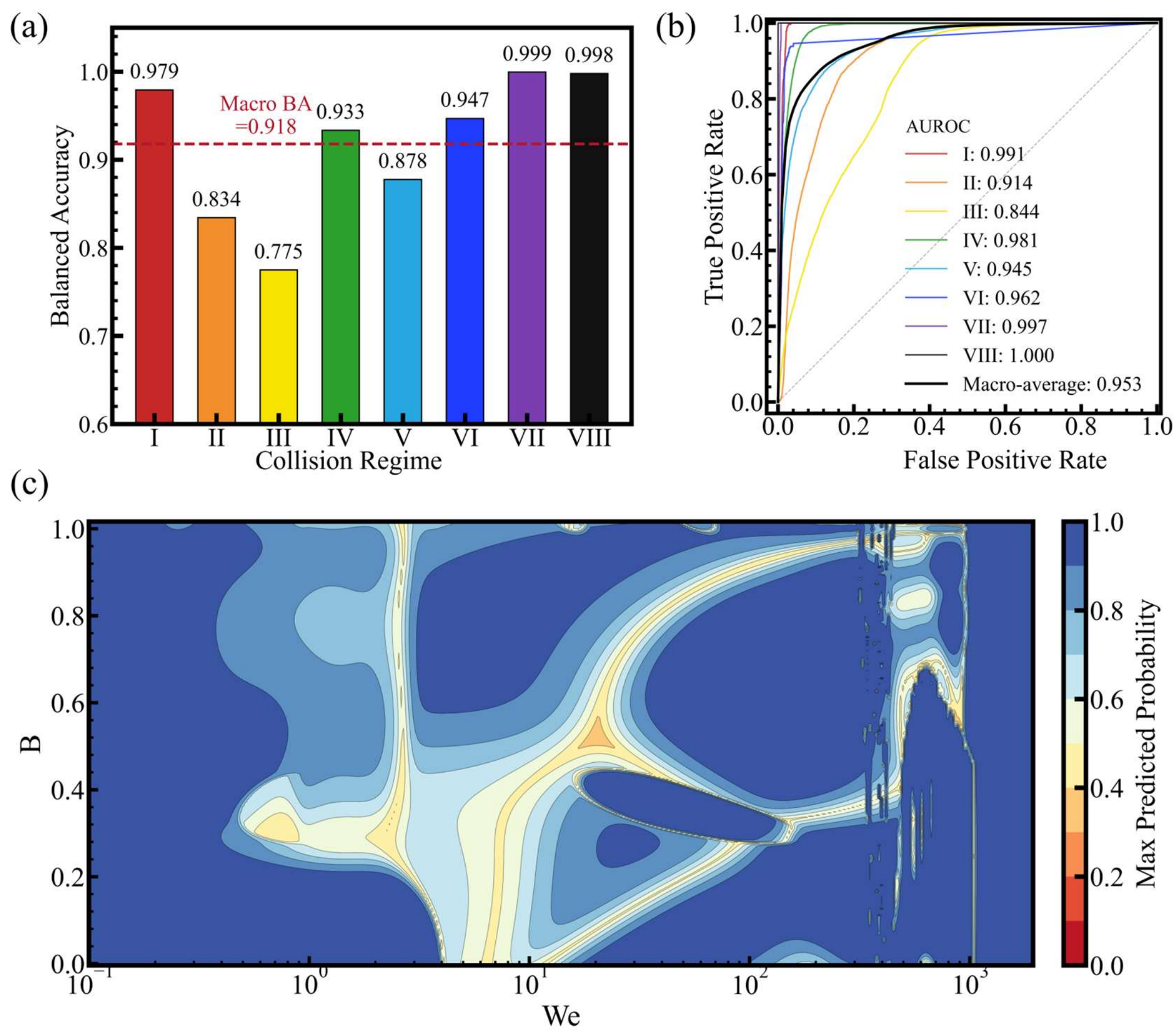

Fig. 6. Predictive performance and fuzzy regime structure of the symbolic-regression model. (a) Balanced accuracy for individual classes and macro-average. (b) ROC curves and corresponding AUROC performance. (c) Fuzzy regime mapping in the We-B plane, obtained by fixing the remaining inputs at $P = 1$, $\Delta = 1$, and $Oh = 0.01$.

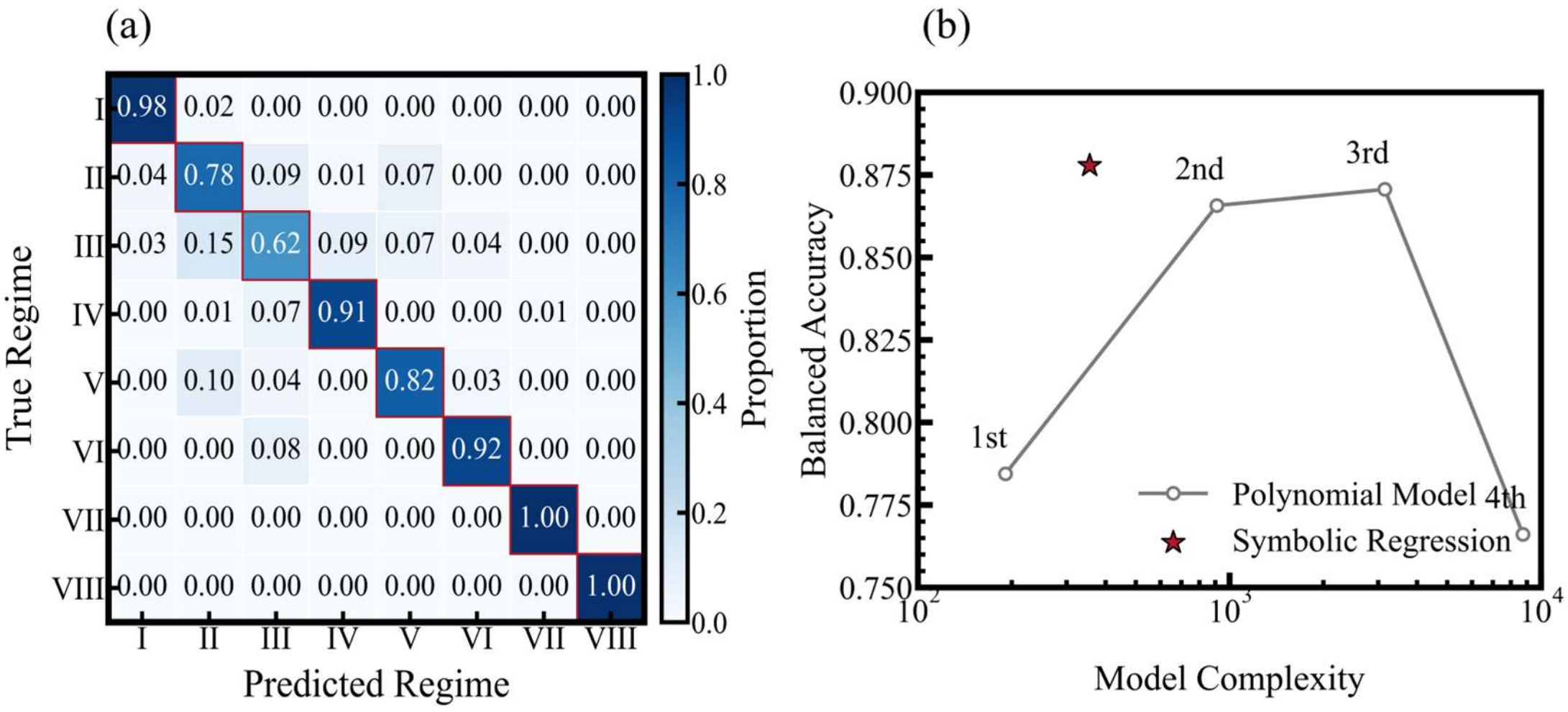

Fig. 7. Classification errors and model-complexity comparison for the symbolic-regression (SR) model. (a) Confusion matrix for classification results. (b) Accuracy–complexity tradeoff between multinomial logistic regression and symbolic regression.

### 3.2 Comparison with Deterministic Analytical Models

The central distinction between the proposed and conventional analytical models lies in their representation of regime transitions. Classical models partition a low-dimensional collision map using prescribed curves and assign each condition to one outcome according to its position relative to the corresponding boundary. The symbolic model instead evaluates a joint probability distribution over all eight outcomes, allowing neighbouring regimes to remain simultaneously plausible where their distributions overlap. As shown in Fig. 8, the classical LB, RS-C, and SS-C criteria pass through the principal transition regions and therefore retain physical value in locating where the dominant outcome changes. The symbolic probability field additionally represents the finite width of these regions and the changing relative likelihoods within them. The distinction is therefore not simply between different transition locations, but between a zero-width deterministic partition and a continuous probability transition.

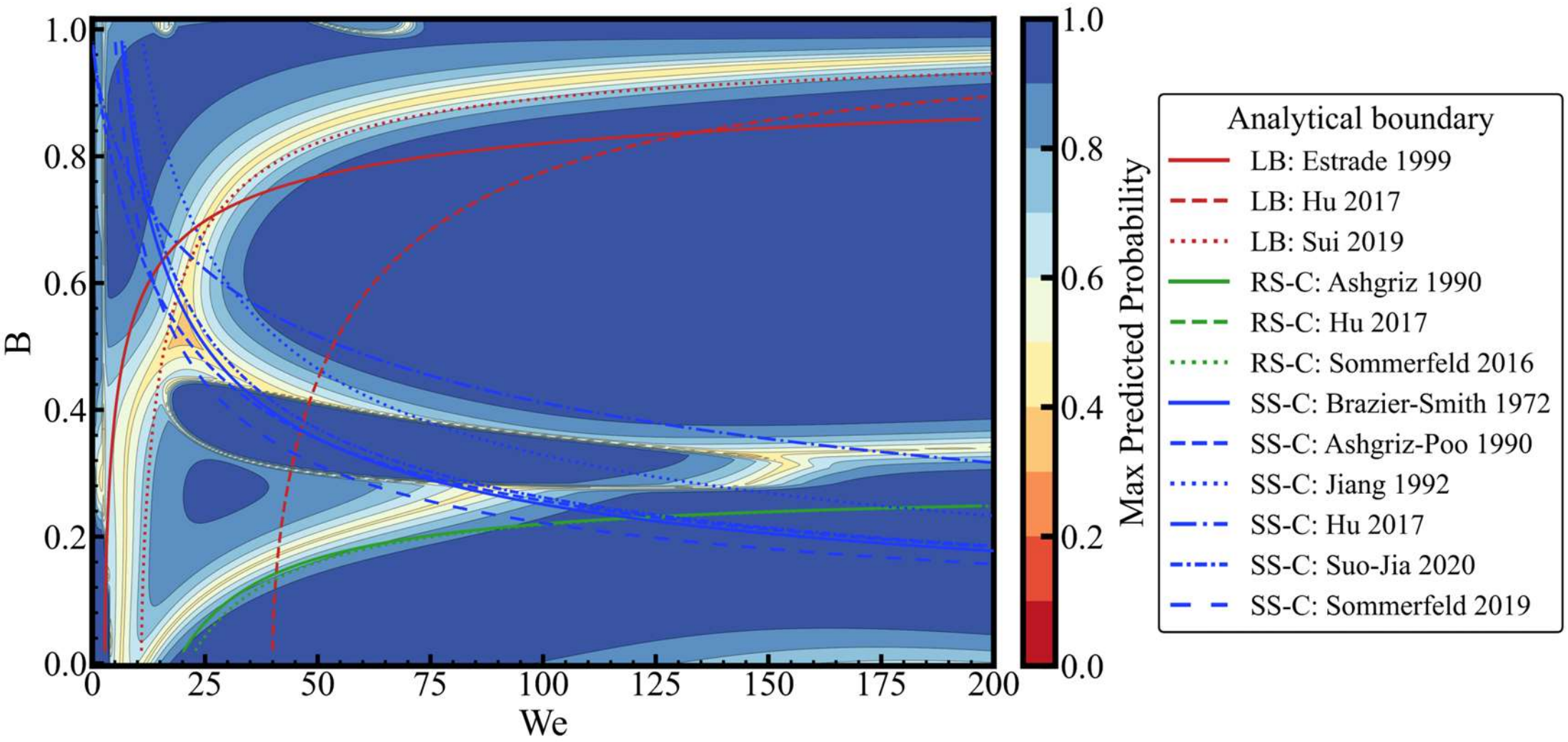

Fig. 8. Symbolic probability field in the $We - B$ plane ($P = 1$, $\Delta = 1$, and $Oh = 0.01$), overlaid with classical analytical boundary models for bouncing-coalescence (LB), reflexive separation-coalescence (RS-C), and stretching separation-coalescence (SS-C) [10, 11, 13, 24, 25, 28, 38, 39].

The finite width of the model-inferred transition regions has a measurable consequence for prediction confidence and regime separability. When regime interiors are defined by $(p_k^{SR})_{max} > 0.9$ and transition bands by $(p_k^{SR})_{max} < 0.7$, the mean top probability decreases from 0.949 to 0.575, while the mean balanced accuracy over regimes II–V decreases from 0.917 to 0.701. This reduced separability does not imply that the classical criteria entirely misplace the transitions. For example, 92% of the reflexive-separation samples in the RS–C band lie within a factor of two of the Ashgriz–Poo boundary [10]. The classical criterion therefore identifies the approximate region in which the outcomes compete, but its zero-width partition cannot represent their overlap within that region.

This distinction is reflected quantitatively in the boundary-resolved comparison in Fig. 9. For LB and SS–C, the symbolic model provides relatively modest improvements over the best analytical criteria, increasing balanced accuracy from 0.786 [28] to 0.834 and from 0.849 [25] to 0.878, respectively. A substantially larger improvement is obtained for RS–C, from 0.620 [13] to 0.933, identifying this strongly overlapping transition as the principal limitation of the deterministic formulation. The ROC curves show the same ordering: the symbolic-model AUROC ranges from 0.914 to 0.981 across the three transitions and reaches 0.981 for RS–C. The agreement between the balanced-accuracy and ROC results shows that the improvement reflects stronger discrimination across thresholds rather than the selection of a favourable operating point.

Figure 10 extends the comparison from isolated transitions to simultaneous prediction of the four canonical regimes. Relative to the multi-boundary Analytical Combined Model (ACM) [21], the balanced-accuracy gains of the symbolic model range from 0.0378 to 0.0835, with the largest improvement obtained for bouncing. By evaluating all candidate outcomes within a common probability space, the symbolic model maintains stronger discrimination when the transitions are considered simultaneously.

Together, Figs. 8–10 distinguish between recalibrating individual transition curves and changing the mathematical representation of collision outcomes. Classical criteria remain informative for locating physically meaningful transitions, but their deterministic structure collapses the finite width and internal competition of these transitions into single switching curves. The symbolic framework retains an explicit analytical form while improving both boundary-resolved and simultaneous multi-regime prediction, with its largest advantages arising where neighbouring outcomes are least separable. The improvement therefore arises from the unified probability representation rather than from constructing another set of recalibrated deterministic boundaries.

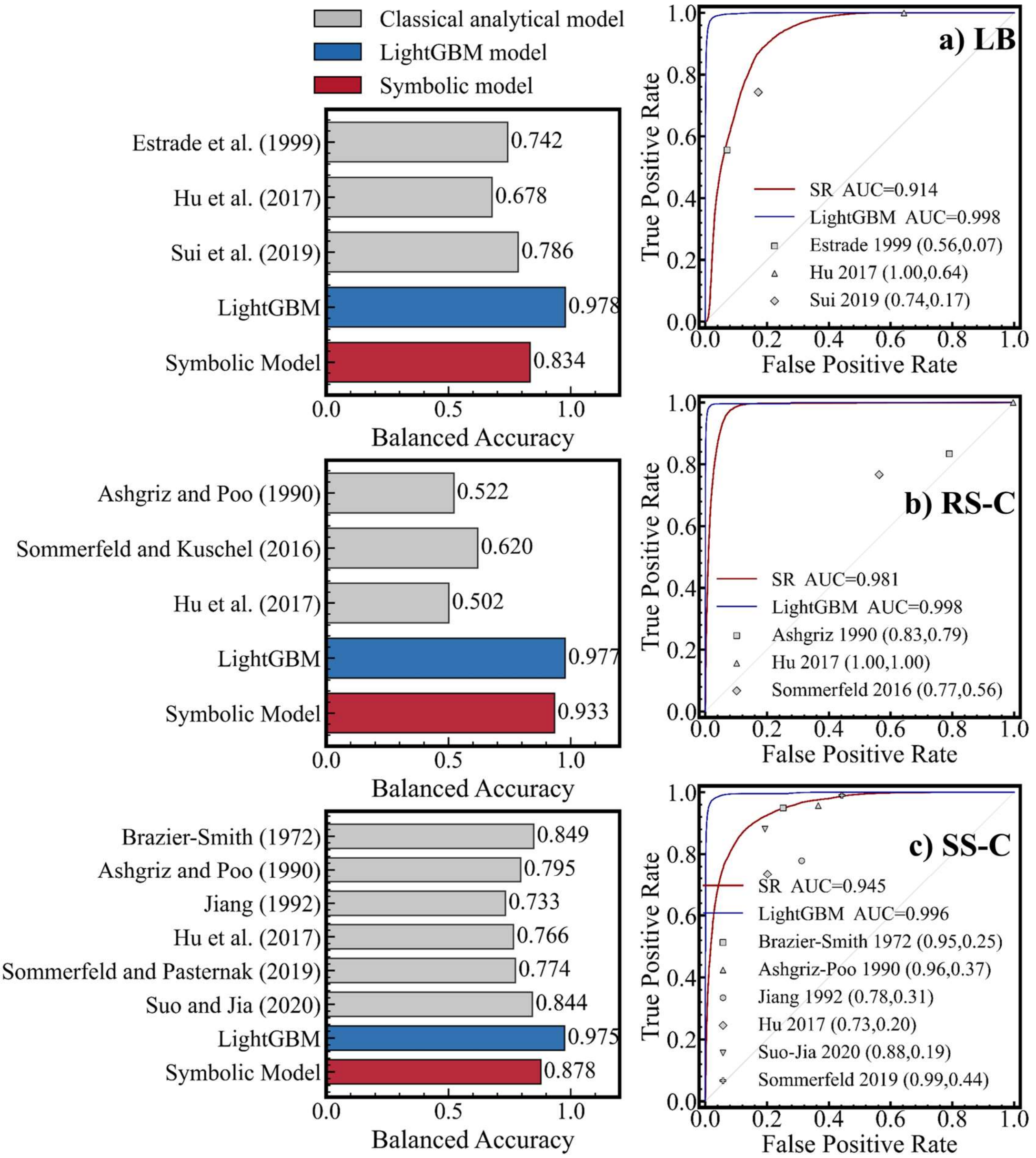

Fig. 9. Boundary-resolved balanced-accuracy and ROC comparison among the classical analytical criteria and the proposed symbolic model for (a) bouncing-coalescence (LB), (b) reflexive-separation-coalescence (RS-C), and (c) stretching-separation-coalescence (SS-C) [10, 11, 13, 24, 25, 28, 38, 39]. The LightGBM teacher is included as a reference.

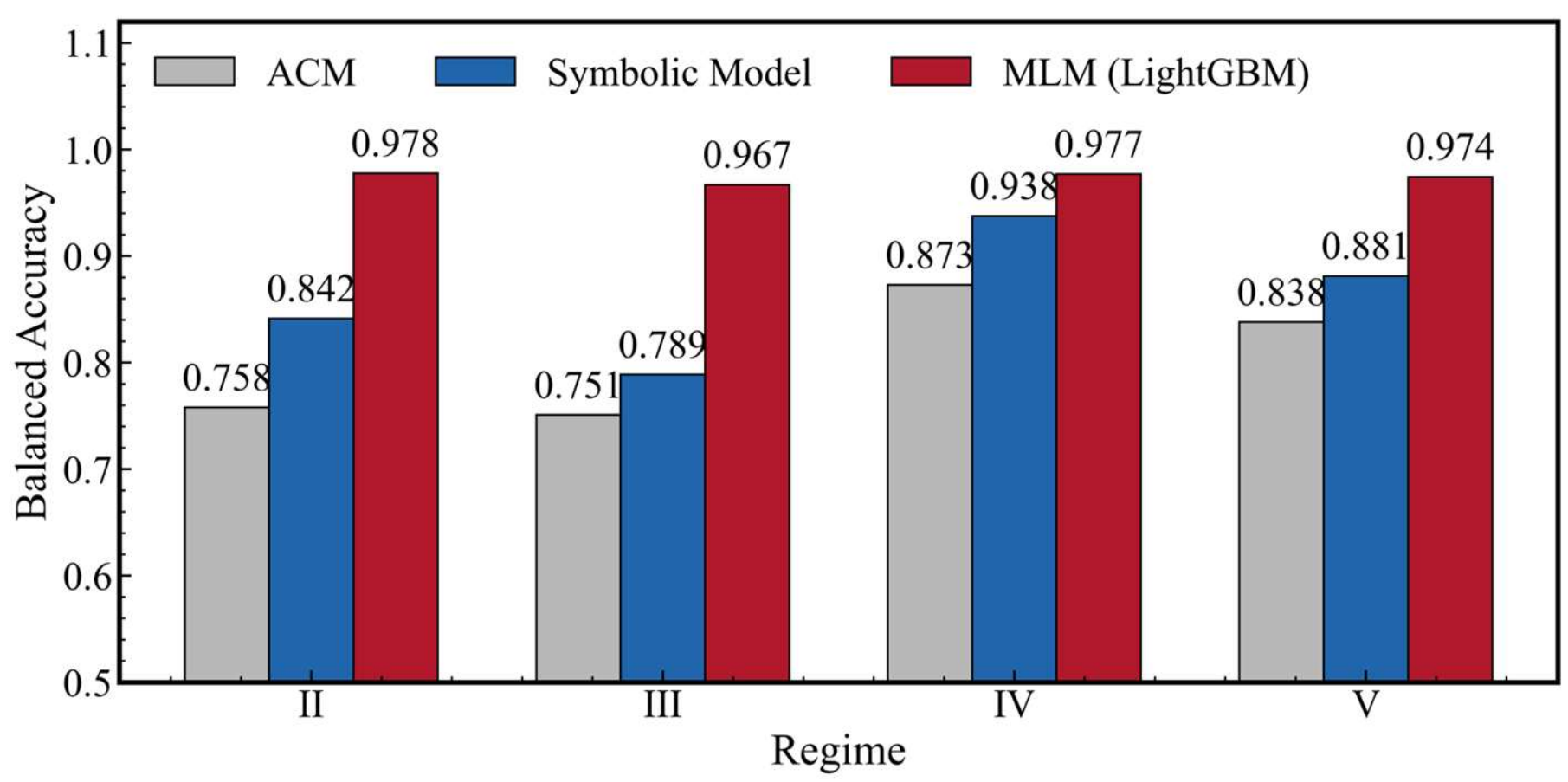

Fig. 10. Balanced accuracy comparison between Analytical Combined Model (ACM) and the proposed symbolic model. The LightGBM teacher is included as a reference.

### 3.3 Stochastic Realization of Collision Outcomes by Multinomial Sampling

Eulerian-Lagrangian spray simulations require a discrete outcome for each collision, whereas the symbolic model returns a probability vector over the eight regimes. For collision $i$, multinomial sampling assigns one categorical outcome $Y_i \sim Cat(\boldsymbol{p}_i^{SR})$, such that $\mathbb{E}[1(Y_i = k)] = \boldsymbol{p}_{i,k}^{SR}$. It therefore preserves the predicted probability mass in expectation, whereas argmax assignment collapses every probability vector to its largest component. This distinction is evident in Fig. 11(a): argmax assignment overrepresents bouncing by 2.83 percentage points and underrepresents rotational separation by 1.70 percentage points relative to the SR-implied probability mass. By contrast, the mean regime fractions over 200 independent realizations reproduce the predicted masses with a maximum absolute deviation of $9 \times 10^{-5}$.

The repeated realizations in Fig. 11(b) quantify the Monte Carlo convergence of this stochastic interface. As the number of draws per event, $M$, increases from 1 to 200, the aggregate sampling error decreases from 0.00493 to 0.000350 and follows the expected $M^{-1/2}$ scaling. This convergence is a numerical verification that the sampling implementation is unbiased and statistically consistent with the symbolic probability field. The practical difference between stochastic and deterministic realization is governed by prediction uncertainty. Agreement between multinomial sampling and argmax assignment decreases from 0.987 in the lowest-entropy interval to below 0.5 when the entropy exceeds approximately 1.23, as shown in Fig. 11(c). Sampling therefore approaches deterministic selection when one outcome dominates but retains alternative outcomes in high-entropy transition regions.

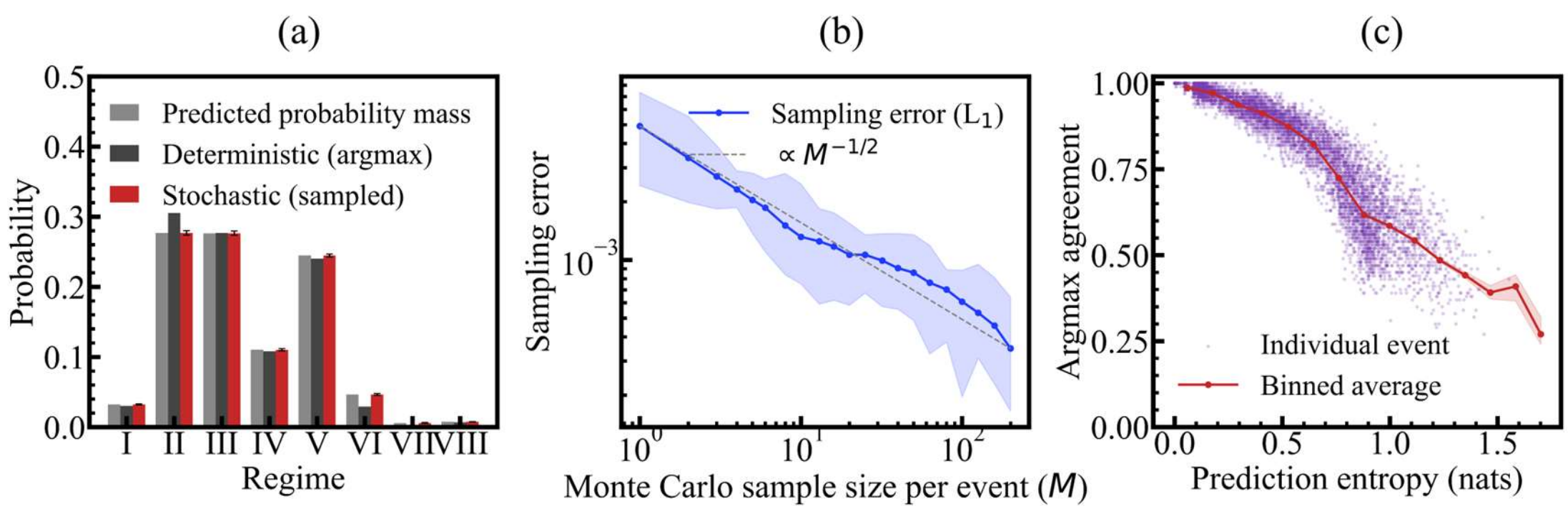

Fig. 11. Statistical behaviour of multinomial collision-outcome sampling: (a) Comparison of the regime fractions obtained from SR probability mass, argmax assignment, and multinomial sampling. (b) Convergence of the aggregate sampling error with the number of realizations $M$. (c) Agreement between multinomial sampling and argmax selection as a function of prediction entropy. Error bars and shaded regions indicate 95% intervals.

Across repeated realizations, the sampled outcomes attain a macro-averaged balanced accuracy of 0.884, with a 95% confidence interval of [0.880,0.887], as shown in Fig. 12. Multinomial sampling thus provides a distribution-preserving and computationally direct interface between the analytical probability model and the discrete outcomes required by Eulerian-Lagrangian simulation. Its influence on spray-scale droplet statistics remains to be assessed in fully coupled simulations.

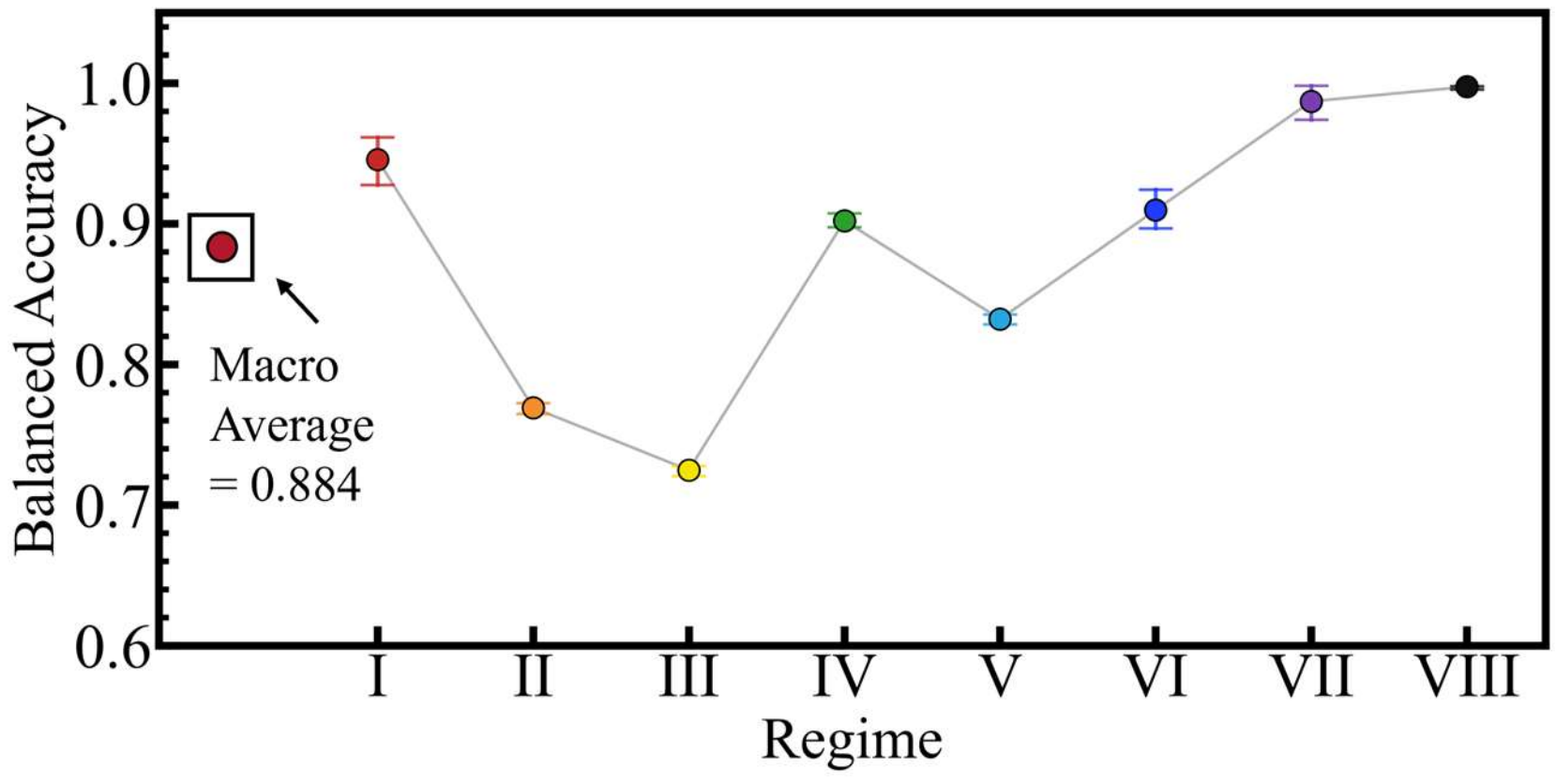

Fig. 12. Balanced accuracy of the sampled outcomes of multinomial sampling: Error bars and shaded regions indicate 95% intervals.

### 3.4 Regime-specific Feature Dependence and Implications for Model Reduction

The finite-width transition regions represented by the symbolic model could arise either from the LightGBM probability landscape or from errors introduced during symbolic compression. Figure 13 distinguishes between these possibilities. The teacher model exhibits reduced maximum probabilities and concentrated misclassifications in the same regions where neighbouring outcomes overlap, while its class-wise performance decreases primarily for the principal transitional regimes (II-V). The spatial correspondence between the teacher and symbolic probability fields therefore shows that symbolic distillation preserves, rather than creates, the transition structure learned by the teacher. This comparison identifies the origin of the transition bands within the modelling pipeline. It does not by itself establish that their finite width represents an intrinsic physical stochasticity of droplet collisions. The regime-resolved SHAP analysis in Fig. 13 and Table 2 shows that no universal feature hierarchy applies to all eight outcomes. Averaged across classes, $B$ and $We$ have the largest mean absolute SHAP values of 2.45 and 2.12, respectively, followed by $Oh$ at 1.44. Their relative importance nevertheless changes substantially among regimes. Within the teacher–student framework, this heterogeneity suggests that different regime-probability functions may benefit from regime-specific reduced input sets rather than a universal descriptor set shared across all outcomes.

Table 2. Mean absolute values of SHAP of each input feature across eight regimes.

| Feature | Regime I | Regime II | Regime III | Regime IV | Regime V | Regime VI | Regime VII | Regime VIII | Average |
|---|---|---|---|---|---|---|---|---|---|
| $\boldsymbol{P}$ | 0.279 | 0.786 | 0.206 | 0.299 | 0.248 | 0.165 | 0.0314 | 0.0433 | 0.257 |
| $\boldsymbol{We}$ | 0.799 | 4.31 | 1.37 | 4.54 | 4.40 | 0.746 | 0.484 | 0.344 | 2.12 |
| $\boldsymbol{B}$ | 0.709 | 3.45 | 3.85 | 3.95 | 4.08 | 1.18 | 0.958 | 1.39 | 2.45 |
| $\boldsymbol{\Delta}$ | 0.590 | 0.309 | 0.743 | 0.592 | 0.25 | 0.108 | 0.0103 | 0.0368 | 0.330 |
| $\boldsymbol{Oh}$ | 1.11 | 2.03 | 0.741 | 4.08 | 1.35 | 1.10 | 0.453 | 0.649 | 1.44 |

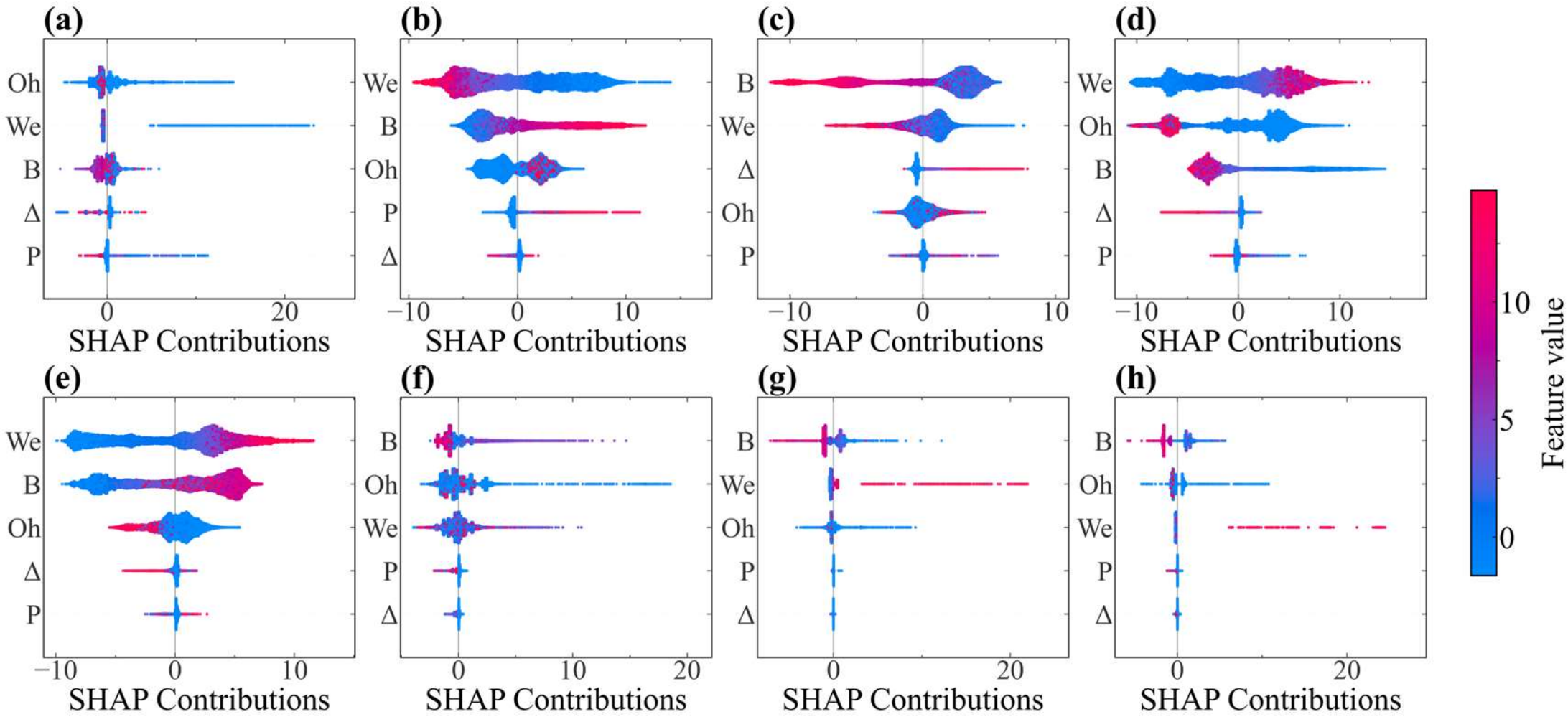

Fig. 13. SHAP analysis revealing feature importance and regime-resolved feature contributions of the LightGBM model. (a) soft coalescence, (b) bouncing, (c) hard coalescence, (d) reflexive separation, (e) stretching separation, (f) rotational separation, (g) finger separation, and (h) splashing.

The regime-dependent SHAP patterns provide a possible route towards reducing the symbolic model as the number of descriptors and collision outcomes increases. The present study retains all five inputs to preserve a common teacher–student representation and does not infer reduced feature sets from SHAP values alone. Future reduction should combine regime-specific feature screening with controlled ablation and additional validation, while imposing physical constraints where a descriptor is known to be essential to a collision mechanism. Such a procedure could limit the rapid growth of the symbolic search space without sacrificing regime discrimination or removing physically necessary dependencies.

## 4. Concluding remarks

This study developed a probabilistic symbolic-distillation model for binary droplet-collision in gaseous environments up to 50 atm. A LightGBM teacher was trained on 38,762 experimental events spanning eight collision regimes and five dimensionless parameters ($We \in [0,2000]$, $Oh \in [9.5 \times 10^{-4}, 5.5 \times 10^{-1}]$, $B \in [0,1]$, $\Delta \in [1,5]$, and $P \in [0.6, 50]$), and its probability landscape was distilled into eight class-specific closed-form expressions. Unlike conventional models that prescribe a separate deterministic curve for each transition between regimes, the present model evaluates the relative likelihoods of all collision outcomes within a single normalised probability space, retaining the flexibility of data-driven learning in an explicit form suitable for Eulerian–Lagrangian spray simulations.

The coupled probability field changes how collision transitions are represented. Instead of imposing zero-width switches, it identifies finite-width fuzzy regions, in which neighbouring outcomes remain simultaneously plausible. The closed-form expressions also recover regime-specific behaviour consistent with established collision physics, including the low-$We$ limit of soft coalescence, the $We - Oh$ combination appearing in the classical splashing parameter, and the low-$B$, near-head-on constraint of reflexive separation, suggesting that the distilled expressions capture physically relevant trends from the data-derived teacher probabilities.

The symbolic-distillation model retains good predictive discrimination across the conditions examined. On the collected eight-regime database, the distilled model achieves a macro-averaged balanced accuracy of 0.918 and a macro AUROC of 0.953, consistently outperforming the conventional deterministic analytical models evaluated in this study. The largest improvement occurs for the strongly overlapping RS-C transition, for which balanced accuracy increases from 0.620 to 0.933. In the simultaneous prediction of the four canonical regimes, the class-wise balanced-accuracy gains over the multi-boundary ACM range from 0.0378 to 0.0835, indicating that individually plausible pairwise boundaries do not necessarily yield a consistent multiclass partition. Finally, multinomial sampling converts the analytical probability vector into a discrete outcome for each collision while preserving the predicted regime distribution at the ensemble level. It therefore provides a distribution-preserving stochastic interface compatible with Eulerian-Lagrangian simulation, avoiding the systematic compositional shifts introduced by deterministic argmax assignment.

The present work establishes and validates the event-level probability model, which achieves a more favourable accuracy and complexity balance than fixed-basis multinomial logistic regression. All five descriptors are retained here to preserve a common teacher–student input space. As the model is extended to higher-dimensional descriptors and finer outcome taxonomies, regime-specific feature screening may also help control the increasing symbolic-search cost. Such reduction should, however, be supported by controlled ablation, physical constraints, and additional validation.

**Acknowledgements.** P.Z. acknowledges support from the National Natural Science Foundation of China (No. 52176134) and partially from the APRC-CityU New Research Initiatives/Infrastructure Support from Central of City University of Hong Kong (No. 9610601). T.Y. thanks Dr. Chenwei Zhang and Dr. Zhenyu Zhang for providing the original experimental data. The authors are grateful to the National Supercomputer Center in Guangzhou (Tianhe-2) for supporting the GPU computing.

**Declaration of interests.** The authors report no conflict of interest.

**Data availability.** The data that support the findings of this study are available from the corresponding author upon reasonable request.

**Supplementary material.** Supporting information of the present model is available.